\documentclass[apjl]{emulateapj}
\usepackage{comment}
\usepackage{ifthen}

\newcommand{\forloop}[5][1]%
{%
\setcounter{#2}{#3}%
\ifthenelse{#4}%
	{%
	#5%
	\addtocounter{#2}{#1}%
	\forloop[#1]{#2}{\value{#2}}{#4}{#5}%
	}%
	{%
	}%
}%

\newcommand{\ctbd}[1]{}

\newcommand{\lc}{light curve}
\newcommand{\lcs}{light curves}
\newcommand{\Lc}{Light curve}

\newcommand{\band}[1]{\ensuremath{#1}~band}

\newcommand{\kms}{\ensuremath{\rm km\,s^{-1}}}
\newcommand{\ms}{\ensuremath{\rm m\,s^{-1}}}

\newcommand{\gcmc}{\ensuremath{\rm g\,cm^{-3}}}
\newcommand{\ergscmsq}{\ensuremath{\rm erg\,s^{-1}\,cm^{-2}}}

\newcommand{\vsini}{\ensuremath{v \sin{i}}}
\newcommand{\feh}{\ensuremath{\rm [Fe/H]}}

\newcommand{\rsun}{\ensuremath{R_\sun}}
\newcommand{\msun}{\ensuremath{M_\sun}}
\newcommand{\lsun}{\ensuremath{L_\sun}}

\newcommand{\rstar}{\ensuremath{R_\star}}
\newcommand{\mstar}{\ensuremath{M_\star}}
\newcommand{\lstar}{\ensuremath{L_\star}}

\newcommand{\teffstar}{\ensuremath{T_{\rm eff\star}}}
\newcommand{\rhostar}{\ensuremath{\rho_\star}}
\newcommand{\loggstar}{\ensuremath{\log{g_{\star}}}}

\newcommand{\rpl}{\ensuremath{R_{p}}}
\newcommand{\mpl}{\ensuremath{M_{p}}}

\newcommand{\rhopl}{\ensuremath{\rho_{p}}}

\newcommand{\arstar}{\ensuremath{a/\rstar}}
\newcommand{\zrstar}{\ensuremath{\zeta/\rstar}}

\newcommand{\rjup}{\ensuremath{R_{\rm J}}}
\newcommand{\mjup}{\ensuremath{M_{\rm J}}}

\newcommand{\refsecl}[1]{\mbox{Section \ref{sec:#1}}}

\newcommand{\reftabl}[1]{Table~\ref{tab:#1}}

\newcommand{\flwof}{\mbox{FLWO 1.2\,m}}

\newcommand{\hatcurCCra}{\ensuremath{17^{\mathrm h}37^{\mathrm m}05.52{\mathrm s}}}                                  
\newcommand{\hatcurCCdec}{\ensuremath{+25{\arcdeg}43{\arcmin}52.3{\arcsec}}}                                 
\newcommand{\hatcurCCtwomass}{2MASS~17370562+2543522}                  
\newcommand{\hatcurCCgsc}{GSC~2080-00517}                              
\newcommand{\hatcurCCtassmv}{\ensuremath{13.207\pm0.039}}              
\newcommand{\hatcurCCtassmB}{\ensuremath{13.871\pm0.039}}              
\newcommand{\hatcurCCtassmI}{\ensuremath{12.67\pm0.14}}                
\newcommand{\hatcurCCtassmg}{\ensuremath{13.50\pm0.04}}            
\newcommand{\hatcurCCtassmr}{\ensuremath{13.06\pm0.02}}            
\newcommand{\hatcurCCtassmi}{\ensuremath{12.88\pm0.04}}            
\newcommand{\hatcurCCtwomassJmag}{\ensuremath{12.020\pm0.022}}         
\newcommand{\hatcurCCtwomassHmag}{\ensuremath{11.714\pm0.026}}         
\newcommand{\hatcurCCtwomassKmag}{\ensuremath{11.627\pm0.025}}         
\newcommand{\hatcurLCrprstar}{\ensuremath{0.1202\pm0.0019}}            
\newcommand{\hatcurLCbsq}{\ensuremath{0.153_{-0.060}^{+0.062}}}        
\newcommand{\hatcurLCimp}{\ensuremath{0.392_{-0.086}^{+0.073}}}        
\newcommand{\hatcurLCzeta}{\ensuremath{18.64\pm0.10}}                  
\newcommand{\hatcurLCdur}{\ensuremath{0.1223\pm0.0013}}                
\newcommand{\hatcurLCingdur}{\ensuremath{0.0152\pm0.0013}}             
\newcommand{\hatcurLCP}{\ensuremath{3.5852467\pm0.0000064}}            
\newcommand{\hatcurLCT}{\ensuremath{2456730.83468\pm0.00027}}          
\newcommand{\hatcurSMEiteff}{\ensuremath{5808\pm50}}                   
\newcommand{\hatcurSMEizfeh}{\ensuremath{-0.030\pm0.080}}              
\newcommand{\hatcurSMEizfehshort}{\ensuremath{-0.03}}                  
\newcommand{\hatcurSMEilogg}{\ensuremath{4.41\pm0.10}}                 
\newcommand{\hatcurSMEivsin}{\ensuremath{1.80\pm0.50}}                 
\newcommand{\hatcurSMEivmac}{\ensuremath{0.0}}                         
\newcommand{\hatcurSMEivmic}{\ensuremath{0.0}}                         
\newcommand{\hatcurSMEiiteff}{\ensuremath{4390\pm50}}                  
\newcommand{\hatcurSMEiizfeh}{\ensuremath{-0.127\pm0.080}}             
\newcommand{\hatcurSMEiizfehshort}{\ensuremath{-0.127}}                
\newcommand{\hatcurSMEiilogg}{\ensuremath{4.66\pm0.10}}                
\newcommand{\hatcurSMEiivsin}{\ensuremath{2.35\pm0.50}}                
\newcommand{\hatcurTRESgamma}{\ensuremath{-9.42\pm0.05}}               
\newcommand{\hatcurLBii}{\ensuremath{0.2619}}                          
\newcommand{\hatcurLBiii}{\ensuremath{0.3313}}                         
\newcommand{\hatcurLBir}{\ensuremath{0.3466}}                          
\newcommand{\hatcurLBiir}{\ensuremath{0.3306}}                         
\newcommand{\hatcurISOmlong}{\ensuremath{1.013\pm0.037}}               
\newcommand{\hatcurISOrlong}{\ensuremath{1.011\pm0.036}}               
\newcommand{\hatcurISOlogg}{\ensuremath{4.434\pm0.032}}                
\newcommand{\hatcurISOlum}{\ensuremath{1.042\pm0.089}}                 
\newcommand{\hatcurISOmv}{\ensuremath{4.785\pm0.097}}                  
\newcommand{\hatcurISOage}{\ensuremath{4.2\pm1.7}}                     
\newcommand{\hatcurISOMK}{\ensuremath{3.266\pm0.081}}                  
\newcommand{\hatcurRVK}{\ensuremath{76.7\pm7.1}}                       
\newcommand{\hatcurRVjitterA}{\ensuremath{0.0\pm3.1}}                  
\newcommand{\hatcurRVjitterB}{\ensuremath{26\pm13}}                    
\newcommand{\hatcurPPi}{\ensuremath{87.70\pm0.56}}                     
\newcommand{\hatcurPPlogg}{\ensuremath{3.012\pm0.060}}                 
\newcommand{\hatcurPPar}{\ensuremath{9.79\pm0.34}}                     
\newcommand{\hatcurPParel}{\ensuremath{0.04604\pm0.00056}}             
\newcommand{\hatcurPPrho}{\ensuremath{0.435\pm0.077}}                  
\newcommand{\hatcurPPm}{\ensuremath{0.582\pm0.056}}                    
\newcommand{\hatcurPPmlong}{\ensuremath{0.582\pm0.056}}                
\newcommand{\hatcurPPr}{\ensuremath{1.182\pm0.055}}                    
\newcommand{\hatcurPPrlong}{\ensuremath{1.182\pm0.055}}                
\newcommand{\hatcurPPmrcorr}{\ensuremath{-0.01}}                       
\newcommand{\hatcurPPteff}{\ensuremath{1313\pm26}}                     
\newcommand{\hatcurPPtheta}{\ensuremath{0.0446\pm0.0048}}              
\newcommand{\hatcurPPfluxavg}{\ensuremath{6.71\pm0.53}}                
\newcommand{\hatcurXAv}{\ensuremath{0.020_{-0.020}^{+0.060}}}          
\newcommand{\hatcurXdistred}{\ensuremath{480\pm19}}                    

\newcommand{\hatcurRVeccentwosiglimemodel}{\ensuremath{<0.139}}

\newcommand{\hatcur}{HAT-P-55}
\newcommand{\hatcurb}{HAT-P-55b}

\newcommand{\hatcurlumind}{\rhostar}
\newcommand{\hatcurjhkfilset}{ESO}

\newcommand{\hatcurSMEversion}{i}                                       
\newcommand{\hatcurSMEteff}{\ifthenelse{\equal{\hatcurSMEversion}{i}}{\hatcurSMEiteff}{\hatcurSMEiiteff}}
\newcommand{\hatcurSMEzfeh}{\ifthenelse{\equal{\hatcurSMEversion}{i}}{\hatcurSMEizfeh}{\hatcurSMEiizfeh}}
\newcommand{\hatcurSMEzfehshort}{\ifthenelse{\equal{\hatcurSMEversion}{i}}{\hatcurSMEizfehshort}{\hatcurSMEiizfehshort}}
\newcommand{\hatcurSMElogg}{\ifthenelse{\equal{\hatcurSMEversion}{i}}{\hatcurSMEilogg}{\hatcurSMEiilogg}}
\newcommand{\hatcurSMEvsin}{\ifthenelse{\equal{\hatcurSMEversion}{i}}{\hatcurSMEivsin}{\hatcurSMEiivsin}}
\newcommand{\hatcurSMEvmac}{\ifthenelse{\equal{\hatcurSMEversion}{i}}{\hatcurSMEivmac}{\hatcurSMEiivmac}}
\newcommand{\hatcurSMEvmic}{\ifthenelse{\equal{\hatcurSMEversion}{i}}{\hatcurSMEivmic}{\hatcurSMEiivmic}}

\newcommand{\hatcurSMEteffcirc}{\ifthenelse{\equal{\hatcurSMEversion}{i}}{\hatcurSMEiteffcirc}{\hatcurSMEiiteffcirc}}
\newcommand{\hatcurSMEzfehcirc}{\ifthenelse{\equal{\hatcurSMEversion}{i}}{\hatcurSMEizfehcirc}{\hatcurSMEiizfehcirc}}
\newcommand{\hatcurSMEzfehshortcirc}{\ifthenelse{\equal{\hatcurSMEversion}{i}}{\hatcurSMEizfehshortcirc}{\hatcurSMEiizfehshortcirc}}
\newcommand{\hatcurSMEloggcirc}{\ifthenelse{\equal{\hatcurSMEversion}{i}}{\hatcurSMEiloggcirc}{\hatcurSMEiiloggcirc}}
\newcommand{\hatcurSMEvsincirc}{\ifthenelse{\equal{\hatcurSMEversion}{i}}{\hatcurSMEivsincirc}{\hatcurSMEiivsincirc}}
\newcommand{\hatcurSMEvmaccirc}{\ifthenelse{\equal{\hatcurSMEversion}{i}}{\hatcurSMEivmaccirc}{\hatcurSMEiivmaccirc}}
\newcommand{\hatcurSMEvmiccirc}{\ifthenelse{\equal{\hatcurSMEversion}{i}}{\hatcurSMEivmiccirc}{\hatcurSMEiivmiccirc}}

\newcommand{\hatcurSMEteffeccen}{\ifthenelse{\equal{\hatcurSMEversion}{i}}{\hatcurSMEiteffeccen}{\hatcurSMEiiteffeccen}}
\newcommand{\hatcurSMEzfeheccen}{\ifthenelse{\equal{\hatcurSMEversion}{i}}{\hatcurSMEizfeheccen}{\hatcurSMEiizfeheccen}}
\newcommand{\hatcurSMEzfehshorteccen}{\ifthenelse{\equal{\hatcurSMEversion}{i}}{\hatcurSMEizfehshorteccen}{\hatcurSMEiizfehshorteccen}}
\newcommand{\hatcurSMEloggeccen}{\ifthenelse{\equal{\hatcurSMEversion}{i}}{\hatcurSMEiloggeccen}{\hatcurSMEiiloggeccen}}
\newcommand{\hatcurSMEvsineccen}{\ifthenelse{\equal{\hatcurSMEversion}{i}}{\hatcurSMEivsineccen}{\hatcurSMEiivsineccen}}
\newcommand{\hatcurSMEvmaceccen}{\ifthenelse{\equal{\hatcurSMEversion}{i}}{\hatcurSMEivmaceccen}{\hatcurSMEiivmaceccen}}
\newcommand{\hatcurSMEvmiceccen}{\ifthenelse{\equal{\hatcurSMEversion}{i}}{\hatcurSMEivmiceccen}{\hatcurSMEiivmiceccen}}

\newcounter{planetcounter}

\newboolean{emulateapj}
\setboolean{emulateapj}{true}

\newboolean{rvtablelong}
\setboolean{rvtablelong}{true}

\newboolean{astroph}
\setboolean{astroph}{true}

\shortauthors{Juncher et al.}
\shorttitle{
\hatcur\lowercase{b}
}
\ifthenelse{\boolean{emulateapj}}{
    \newcommand{\titledag}{$\dagger$}
}{
    \newcommand{\titledag}{\dagger}
}

\begin{document}
\title{
\hatcur\lowercase{b}: A Hot Jupiter Transiting a Sun-Like Star
\altaffilmark{\titledag}
}

\author{
    D.~Juncher\altaffilmark{1,2}, L.~A.~Buchhave\altaffilmark{2,3}, J.~D.~Hartman\altaffilmark{4}, G.~\'A.~Bakos\altaffilmark{4}, A.~Bieryla\altaffilmark{3}, 
    T.~Kov\'acs\altaffilmark{4,5,8}, I.~Boisse\altaffilmark{6}, D.~W.~Latham\altaffilmark{3}, G.~Kov\'acs\altaffilmark{5}, W.~Bhatti\altaffilmark{4}, 
    Z.~Csubry\altaffilmark{4}, K.~Penev\altaffilmark{4}, M.~de~Val-Borro\altaffilmark{4}, E.~Falco\altaffilmark{3}, 
    G.~Torres\altaffilmark{3}, R.~W.~Noyes\altaffilmark{3}, J.~L\'az\'ar\altaffilmark{7}, I.~Papp\altaffilmark{7}, P.~S\'ari\altaffilmark{7}   
}

\altaffiltext{1}{Niels Bohr Institute, University of Copenhagen, DK-2100, Denmark}
\altaffiltext{2}{Centre for Star and Planet Formation, Natural History Museum of Denmark, University of Copenhagen, DK-1350 Copenhagen}
\altaffiltext{3}{Harvard-Smithsonian Center for Astrophysics, Cambridge, MA 02138 USA}
\altaffiltext{4}{Department of Astrophysical Sciences, Princeton University, Princeton, NJ 08544 USA}
\altaffiltext{5}{Konkoly Observatory, Budapest, Hungary}
\altaffiltext{6}{Aix Marseille Universit\'e, CNRS, LAM (Laboratoire d'Astrophysique de Marseille) UMR 7326, 13388, Marseille, France}
\altaffiltext{7}{Hungarian Astronomical Association (HAA)}
\altaffiltext{8}{Fulbright Fellow}

\altaffiltext{$\dagger$}{
Based on observations obtained with the Hungarian-made Automated
Telescope Network. Based in part on radial velocities obtained with
the SOPHIE spectrograph mounted on the 1.93\,m telescope at
Observatoire de Haute-Provence, France. Based in part on observations
made with the Nordic Optical Telescope, operated on the island of La
Palma jointly by Denmark, Finland, Iceland, Norway, and Sweden, in the
Spanish Observatorio del Roque de los Muchachos of the Instituto de
Astrof\'isica de Canarias. Based in part on observations obtained with
the Tillinghast Reflector 1.5\,m telescope and the 1.2\,m telescope,
both operated by the Smithsonian Astrophysical Observatory at the Fred
Lawrence Whipple Observatory in Arizona.
}


\begin{abstract}

\setcounter{footnote}{10}
We report the discovery of a new transiting extrasolar planet, \hatcurb{}. 
The planet orbits a {V = \hatcurCCtassmv\ }sun-like star with a mass of 
\hatcurISOmlong\ \msun, a radius of \hatcurISOrlong\ \rsun\ 
and a metallicity of $-0.03$ $\pm$ $0.08$. The planet itself is a typical hot Jupiter 
with a period of \hatcurLCP\ days, a mass of \hatcurPPm\ \mjup\ 
and a radius of \hatcurPPr\ \rjup. This discovery adds to the
increasing sample of transiting planets with measured bulk densities, which 
is needed to put constraints on models of planetary structure and formation theories.
\setcounter{footnote}{0}
\end{abstract}


\section{Introduction}
\label{sec:introduction}
Today we know of almost 1800 validated exoplanets and more than 4000 exoplanet 
candidates. Among these, the transiting exoplanets (TEPs) are essential in our 
exploration and understanding of the physical properties of exoplanets. While radial
velocity observations alone only allow us to estimate the minimum mass of a planet, we 
can combine them with transit observations for a more comprehensive study of the physical 
properties of the planet. 
When a planet transits, we can determine the inclination of the orbit and the radius of the
planet, allowing us to break the mass degeneracy and, along with the mass, determine
the mean density of the planet. The mean density of a planet offers us insight into its 
interior composition, and although there is an inherent degeneracy 
arising from the fact that planets of different compositions can have identical masses
and radii, this information allows us to map the diversity and distribution of exoplanets
and even put constraints on models of planetary structure and formation theories.

The occurrence rate of hot Jupiters in the Solar neighbourhood is around 1\% 
(\citealt{udrysantos:2007}; \citealt{wright:2012}). With a transit probability
of about $\sim$10$\%$, roughly one thousand stars need to be monitored in 
photometry to find just a single hot Jupiter. Therefore, the majority of the known 
transiting hot Jupiters have been discovered by photometric wide field surveys 
targeting tens of thousands of stars per night. 

In this paper we present the discovery of \hatcurb{} by the Hungarian-made 
Automated Telescope Network (HATNet; see \citealt{bakos:2004:hatnet}), a 
network of six small fully automated wide field telescopes of which four 
are located at the Fred Lawrence Whipple Observatory in Arizona,
and two are located at the Mauna Kea Observatory in Hawaii. Since HATNet 
saw first light in 2003, it has searched for TEPs around bright stars (V $\lesssim$ 13) 
covering about 37\% of the Northern sky, and discovered approximately 25\%
of the known transiting hot Jupiters. 

The layout of the paper is as follows. In Section 2 we present the different
photometric and spectroscopic observations that lead to the detection and
characterisation of \hatcurb{}. In Section 3 we derive the stellar and planetary
parameters. Finally, we discuss the characteristics of \hatcurb{} in Section 4. 

\section{Observations}
\label{sec:obs}
The general observational procedure used by HATNet to
discover TEPs has been described in detail in previous
papers (e.g. \citealt{bakos:2010:hat11}; \citealt{latham:2009:hat8}).
In this section we present the specific details for the discovery and
follow-up observations of \hatcurb{}.

\subsection{Photometry}
\label{sec:photometry}

\subsubsection{Photometric detection}
\label{sec:detection}
\hatcurb{}  was initially identified as a candidate transiting exoplanet 
based on photometric observations made by the HATNet survey 
\citep{bakos:2004:hatnet} in 2011. The observations of 
\hatcur{} were made on nights between February and August  
with the HAT-5 telescope at the Fred Lawrence Whipple 
Observatory (FLWO) in Arizona, and on nights between May and August 
with the HAT-8 telescope at Mauna Kea Observatory in Hawaii. 
Both telescopes used a Sloan \band{r} filter. HAT-5 provided a total of 
10574 images with a median cadence of 218s, and HAT-8 provided a 
total of 6428 images with a median cadence of 216s. 

\begin{figure}[h]
\plotone{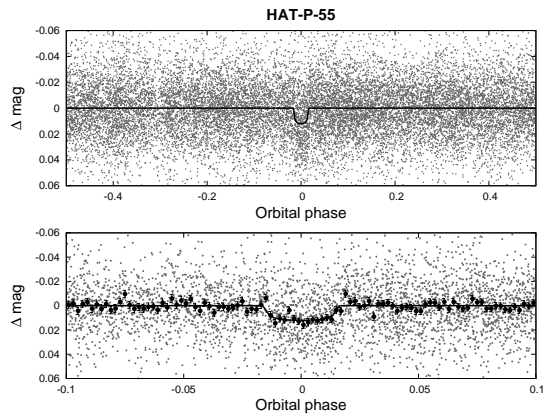}
\caption[]{
    HATNet \lc{} of \hatcur\ phase folded with the transit period.
    The top panel shows the unbinned light curve,
    while the bottom panel shows the region zoomed-in on the transit, with
    dark filled circles for the light curve binned in phase with a
    binsize of 0.002. The solid line represents the best-fit light
    curve model.
\label{fig:hatnet}}
\end{figure}

The results were processed and reduced to trend-filtered light curves using 
the External Parameter Decorrelation method (EPD; see \citealt{bakos:2010:hat11}) 
and the Trend Filtering Algorithm (TFA; see \citealt{kovacs:2005:TFA}).
The light curves were searched for periodic transit signals using the Box
Least-Squares method (BLS; see \citealt{kovacs:2002:BLS}). The individual 
photometric measurements for \hatcur{} are listed in Table~\ref{tab:phfu}, 
and the folded light curves together with the best-fit transit light curve model 
are presented in Figure~\ref{fig:hatnet}.

\subsubsection{Photometric follow-up}
\label{sec:phfu}
We performed photometric follow-up observations of \hatcur{}  
using the KeplerCam CCD camera on the {1.2 m} telescope at 
the FLWO, observing a transit ingress on the night of 23 May 2013, 
and a full transit on the night of 7 April 2014. Both transits were
observed using a Sloan $i$-band filter. For the first event 
we obtained 230 images with a median cadence of 64s, and 
for the second event we obtained 258 images with a median 
cadence of 67s. 

\begin{figure}[]
\plotone{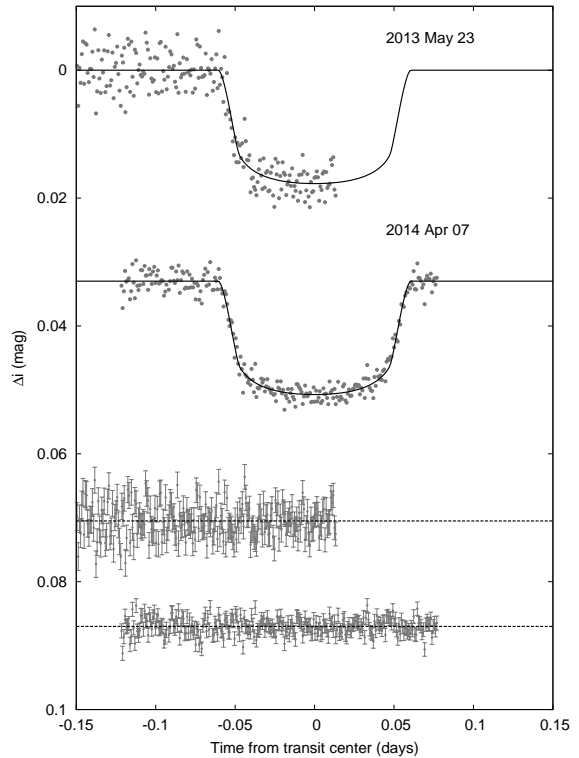}
\caption{
    Unbinned transit \lcs{} for \hatcur, acquired with KeplerCam at
    the \flwof{} telescope. The light curves have been EPD- and
    TFA-processed, as described in \citet{bakos:2010:hat11}.  
    The solid lines represents the best fit from the global
    modeling described in \refsecl{analysis}. Residuals from the fit 
    are displayed below at the bottom of the figure. The error bars 
    represent the photon and background shot noise, plus the readout 
    noise.
}
\label{fig:lc}
\end{figure}

The results were reduced to light curves following the procedure of 
\citet{bakos:2010:hat11}, and EPD and TFA were performed to
remove trends simultaneously with the light curve modelling. The 
individual photometric follow-up measurements for \hatcur{} 
are listed in Table~\ref{tab:phfu}, and the folded light curves 
together with our best-fit transit light curve model are presented 
in Figure~\ref{fig:lc}.

Subtracting the transit signal from the HATNet light curve, we 
used the BLS method to search for additional transit signals and 
found none. A Discrete Fourier Transform also revealed no other 
periodic signals in the data. 

\ifthenelse{\boolean{emulateapj}}{
    \begin{deluxetable*}{lrrrrr}
}{
    \begin{deluxetable}{lrrrrr}
}
\tablewidth{0pc}
\tablecaption{
    Differential photometry of
    \hatcur\label{tab:phfu}.
}
\tablehead{
    \colhead{BJD\tablenotemark{a}} & 
    \colhead{Mag\tablenotemark{b}} & 
    \colhead{\ensuremath{\sigma_{\rm Mag}}} &
    \colhead{Mag(orig)\tablenotemark{c}} & 
    \colhead{Filter} &
    \colhead{Instrument} \\
    \colhead{\hbox{~~~~(2,400,000$+$)~~~~}} & 
    \colhead{} & 
    \colhead{} &
    \colhead{} & 
    \colhead{} & 
    \colhead{}
}
\startdata
$ 55667.80829 $ & $  -0.01089 $ & $   0.02455 $ & $ \cdots $ & $ r$ &     HATNet\\
$ 55649.88231 $ & $   0.00947 $ & $   0.01537 $ & $ \cdots $ & $ r$ &     HATNet\\
$ 55778.95165 $ & $   0.02246 $ & $   0.01538 $ & $ \cdots $ & $ r$ &     HATNet\\
$ 55692.90580 $ & $  -0.00422 $ & $   0.01713 $ & $ \cdots $ & $ r$ &     HATNet\\
$ 55735.92912 $ & $   0.00602 $ & $   0.02207 $ & $ \cdots $ & $ r$ &     HATNet\\
$ 55771.78174 $ & $  -0.01054 $ & $   0.01241 $ & $ \cdots $ & $ r$ &     HATNet\\
$ 55710.83270 $ & $  -0.01741 $ & $   0.01609 $ & $ \cdots $ & $ r$ &     HATNet\\
$ 55674.98046 $ & $   0.01418 $ & $   0.01771 $ & $ \cdots $ & $ r$ &     HATNet\\
$ 55728.75933 $ & $   0.04136 $ & $   0.03816 $ & $ \cdots $ & $ r$ &     HATNet\\
$ 55710.83315 $ & $   0.00822 $ & $   0.01348 $ & $ \cdots $ & $ r$ &     HATNet\\

\enddata
\tablenotetext{a}{
    Barycentric Julian Date calculated directly from UTC, {\em
      without} correction for leap seconds.
}
\tablenotetext{b}{
    The out-of-transit level has been subtracted. These magnitudes
    have been subjected to the EPD and TFA procedures, carried out
    simultaneously with the transit fit for the follow-up data. For
    HATNet this filtering was applied {\em before} fitting for the
    transit.
}
\tablenotetext{c}{
    Raw magnitude values after correction using comparison stars, but
    without application of the EPD and TFA procedures. This is only
    reported for the follow-up light curves.\\\\\\\\
}
\tablecomments{
    This table is available in a machine-readable form in the online
    journal.  A portion is shown here for guidance regarding its form
    and content.
}
\ifthenelse{\boolean{emulateapj}}{
    \end{deluxetable*}
}{
    \end{deluxetable}
}

\ifthenelse{\boolean{emulateapj}}{
    \begin{deluxetable*}{lrrrrrr}
}{
    \begin{deluxetable}{lrrrrrr}
}
\tablewidth{0pc}
\tablecaption{
    Relative radial velocities, and bisector span measurements of \hatcur.
    \label{tab:rvs}
}
\tablehead{
    \colhead{BJD\tablenotemark{a}} &
    \colhead{RV\tablenotemark{b}} &
    \colhead{\ensuremath{\sigma_{\rm RV}}\tablenotemark{c}} &
    \colhead{BS} &
    \colhead{\ensuremath{\sigma_{\rm BS}}} &
    \colhead{Phase} &
    \colhead{Instrument}\\
    \colhead{\hbox{(2,456,000$+$)}} &
    \colhead{(\ms)} &
    \colhead{(\ms)} &
    \colhead{(\ms)} &
    \colhead{(\ms)} &
    \colhead{} &
    \colhead{}
}
\startdata
$ 427.61363 $ & $   -50.8 $ & $    19.0 $ & $    20.0 $ & $    38.0 $ & $   0.425 $ &  FIES \\
$ 428.54873 $ & $    90.9 $ & $    16.0 $ & $    -5.0 $ & $    32.0 $ & $   0.686 $ &  FIES \\
$ 446.59498 $ & $   102.0 $ & $    15.1 $ & $   -15.9 $ & $    30.2 $ & $   0.720 $ &  OHP \\
$ 448.51197 $ & $   -39.8 $ & $    13.1 $ & $     7.5 $ & $    26.2 $ & $   0.254 $ &  OHP \\
$ 450.47096 $ & $    61.0 $ & $     9.1 $ & $    -8.2 $ & $    18.2 $ & $   0.801 $ &  OHP \\
$ 454.49007 $ & $     9.8 $ & $     9.1 $ & $    -2.2 $ & $    18.2 $ & $   0.922 $ &  OHP \\
$ 455.52529 $ & $   -86.4 $ & $    10.9 $ & $     3.4 $ & $    21.8 $ & $   0.211 $ &  OHP \\
$ 456.44155 $ & $   -48.4 $ & $     6.8 $ & $     2.2 $ & $    13.6 $ & $   0.466 $ &  OHP \\
$ 528.43752 $ & $    43.8 $ & $    17.7 $ & $   -50.0 $ & $    35.4 $ & $   0.547 $ &  FIES \\
$ 528.48816 $ & $    18.5 $ & $    15.6 $ & $   -31.0 $ & $    31.2 $ & $   0.561 $ &  FIES \\
$ 529.50540 $ & $    54.6 $ & $    14.7 $ & $     0.0 $ & $    29.4 $ & $   0.845 $ &  FIES \\
$ 529.55576 $ & $    38.8 $ & $    23.8 $ & $   -27.0 $ & $    47.6 $ & $   0.859 $ &  FIES \\
$ 530.41408 $ & $   -26.2 $ & $    13.0 $ & $    22.0 $ & $    26.0 $ & $   0.099 $ &  FIES \\
$ 530.46398 $ & $   -51.0 $ & $    15.6 $ & $    18.0 $ & $    31.2 $ & $   0.112 $ &  FIES \\
$ 531.47003 $ & $   -51.4 $ & $    14.4 $ & $    24.0 $ & $    28.8 $ & $   0.393 $ &  FIES \\
$ 531.52009 $ & $   -57.7 $ & $    13.0 $ & $    27.0 $ & $    26.0 $ & $   0.407 $ &  FIES \\

\enddata
\tablenotetext{a}{
    Barycentric Julian Date calculated directly from UTC, {\em
      without} correction for leap seconds.
}
\tablenotetext{b}{
    The zero-point of these velocities is arbitrary. An overall offset
    $\gamma_{\rm rel}$ fitted to these velocities in \refsecl{analysis}
    has {\em not} been subtracted. 
}
\tablenotetext{c}{
    Internal errors excluding the component of astrophysical jitter
    considered in \refsecl{analysis}.
}
\ifthenelse{\boolean{rvtablelong}}{
}{
} 
\ifthenelse{\boolean{emulateapj}}{
    \end{deluxetable*}
}{
    \end{deluxetable}
}

\subsection{Spectroscopy}
\label{sec:hispec}
We performed spectroscopic follow-up observations of \hatcur{} 
to rule out false positives and to determine the RV variations
and stellar parameters. Initial reconnaissance observations were 
carried out with the Tillinghast Reflector Echelle Spectrograph (TRES; 
\citealt{furesz:2008}) at the FLWO. We obtained 2 spectra near opposite 
quadratures on the nights of 4 and 31 October 2012. Using the Stellar Parameters Classification 
method (SPC; see \citealt{buchhave:2012}), we determined the initial RV measurements 
and stellar parameters. We found a mean absolute RV of $-9.42$ km s$^{-1}$ 
with an rms of 48 m s$^{-1}$, which is consistent with no detectable
RV variation. The stellar parameters, including the effective temperature
\teffstar = 5800 $\pm $ 50 K, surface gravity \loggstar = 4.5 $\pm $ 0.1 
(log cgs) and projected rotational velocity \vsini = 5.0 $\pm $ 0.4 km s$^{-1}$,
correspond to those of a G2 dwarf.

High-resolution spectroscopic observations were then carried out 
with the SOPHIE spectrograph mounted on the 1.93 m 
telescope at Observatoire de Haute-Provence (OHP) 
\citep{perruchot:2011, bouchy:2013}, and with the FIES spectrograph 
mounted on the 2.6 m Nordic Optical Telescope \citep{djupvik:2010}. We 
obtained 6 SOPHIE spectra on nights between 3 June and 12 June 2013, 
and 10 FIES spectra on nights between 15 May and 26 August 2013.

\begin{figure} [h]
\plotone{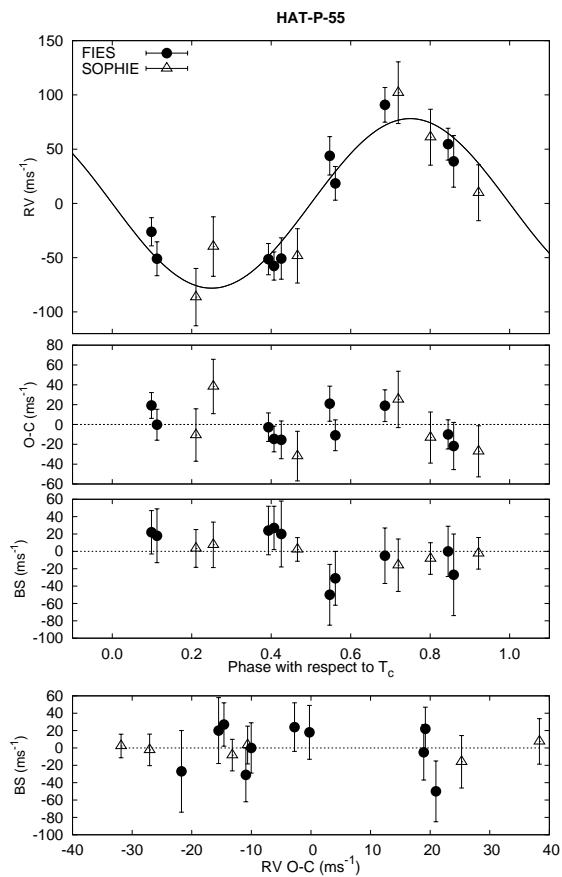}
\caption{
    {\em Top panel:} RV measurements from NOT~2.6\,m/FIES (filled
    circles) and OHP~1.93\,m/SOPHIE (open triangles) for
    \hbox{\hatcur} shown as a function of orbital phase, along with
    our best-fit circular model (solid line; see
    \reftabl{planetparam}).  Zero phase corresponds to the time of
    mid-transit.  The center-of-mass velocity has been subtracted.
    {\em Second panel:} RV residuals from our best-fit circular
    model. The error bars include a ``jitter'' component
    (\hatcurRVjitterA\,\ms, and \hatcurRVjitterB\,\ms\ for FIES and
    SOPHIE respectively) added in quadrature to the formal errors (see
    \refsecl{hispec}). The symbols are as in the upper panel.
    {\em Third panel:} Bisector spans (BS), adjusted to have a median of 0.
    {\em Bottom panel:} RV residuals from our best-fit circular model vs. BS. 
    There is no sign of correlation.  
    Note the different vertical scales of the panels.
}
\label{fig:rvbis}
\end{figure}

We reduced and extracted the spectra and derived radial velocities 
and spectral line bisector span (BS) measurements following the 
method of \citet{boisse:2012:hat4243} for the SOPHIE data and the 
method of \citet{buchhave:2010:hat16}  for the FIES data. The final RV 
data and their 
errors are listed for both instruments in Table~\ref{tab:rvs}, and 
\newpage
\mbox{}
\newpage
\mbox{}
\newpage
\noindent
the folded 
RV data together with our best-fit orbit and corresponding residuals and 
bisectors are presented in Figure~\ref{fig:rvbis}. To avoid underestimating the
BS uncertainties we base them directly on the RV uncertainties, setting them
equal to twice the RV uncertainties. At a first glance there do seem to be a 
slight hint of variation of the BSs in phase with the RVs suggesting there
might be a blend. This is not the case, however, as will we show in our detailed blend
analysis in Section 3.

We applied the SPC method to the FIES spectra to
determine the final spectroscopic parameters of \hatcur{}. The 
values were calculated using a weighted mean, taking into account
the cross correlation function (CCF) peak height. The results are 
shown in Table~\ref{tab:stellar}.

\setcounter{planetcounter}{1}
%

\section{Analysis}
\label{sec:analysis}
In order to rule out the possibility that \hatcur{} is a blended stellar 
eclipsing binary system, and not a transiting planet system, we 
carried out a blend analysis following \cite{hartman:2012:hat39hat41}.  
We find that a single star with a transiting planet fits the light curves 
and catalog photometry better than models involving a stellar eclipsing 
binary blended with light from a third star. While it is possible to marginally 
fit the photometry using a G star eclipsed by a late M dwarf that is blended 
with another bright G star, simulated spectra for this scenario are obviously 
composite and show large (multiple \kms) bisector span and RV variations 
that are inconsistent with the observations. Based on this analysis we 
conclude that \hatcur{} is not a blended stellar eclipsing binary system, and 
is instead best explained as a transiting planet system. We also consider 
the possibility that \hatcur{} is a planetary system with a low-mass stellar 
companion that has not been spatially resolved. The constraint on this 
scenario comes from the catalog photometric measurements, based on 
which we can exclude a physical companion star with a mass greater 
than $0.7$\,\msun. Any companion would dilute both the photometric transit 
and radial velocity orbit. The maximum dilution allowed by the photometry 
would increase the planetary radius by $\sim$15\%.

\begin{figure}[h]
\plotone{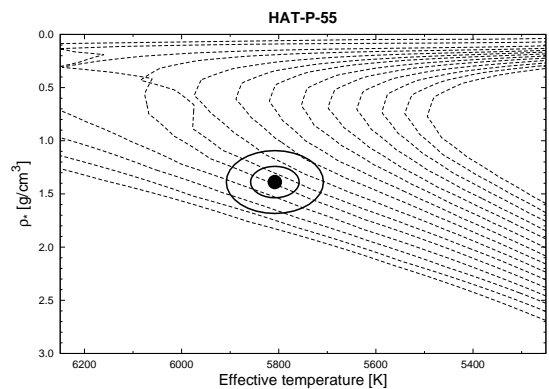}
\caption[]{
    Comparison between the measured values of \teffstar\ and
    \rhostar\ (from SPC applied to the FIES spectra, and from our
    modeling of the light curves and RV data, respectively), and the
    Y$^{2}$ model isochrones from \citet{yi:2001}. The best-fit
    values, and approximate 1$\sigma$ and 2$\sigma$ confidence
    ellipsoids are shown. The Y$^{2}$ isochrones are shown for ages of
    0.2\,Gyr, and 1.0 to 14.0\,Gyr in 1\,Gyr increments.
\label{fig:iso}}
\end{figure}

We analyzed the system following the procedure of \citet{bakos:2010:hat11}
with modifications described in \cite{hartman:2012:hat39hat41}. In short, we 
(1) determined the stellar atmospheric parameters of the
host star \hatcur{} by applying the SPC method to the FIES 
spectra; (2) used a Differential Evolution Markov-Chain Monte Carlo
procedure to simultaneously model the RVs and the light curves, keeping 
the limb darkening coefficients fixed to those of the \cite{claret:2004} tabulations; 
(3) used the spectroscopically inferred effective temperatures
and metallicities of the star, the stellar densities determined from the light
curve modeling, and the Yonsei-Yale theoretical stellar evolution models 
\citep{yi:2001} to determine the stellar mass, radius and age, as well as
the planetary parameters (e.g. mass and radius) which depend on the
stellar values (Figure~\ref{fig:iso}).

\ifthenelse{\boolean{emulateapj}}{
  \begin{deluxetable*}{lcr}
}{
  \begin{deluxetable}{lcr}
}
\tablewidth{0pc}
\tabletypesize{\scriptsize}
\tablecaption{
    Stellar Parameters for \hatcur{} 
    \label{tab:stellar}
}
\tablehead{
    \multicolumn{1}{c}{~~~~~~~~Parameter~~~~~~~~} &
    \multicolumn{1}{c}{Value}                     &
    \multicolumn{1}{c}{Source}    
}
\startdata
\noalign{\vskip -3pt}
\sidehead{Identifying Information}
~~~~R.A. (h:m:s)                      &  \hatcurCCra{} & 2MASS\\
~~~~Dec. (d:m:s)                      &  \hatcurCCdec{} & 2MASS\\
~~~~GSC ID                            &  \hatcurCCgsc{} & GSC\\
~~~~2MASS ID                          &  \hatcurCCtwomass{} & 2MASS\\
~~~~HTR ID                            & HTR 287-004 & HATNet\\
\sidehead{Spectroscopic properties}
~~~~$\teffstar$ (K)\dotfill         &  \hatcurSMEteff{} & SPC \tablenotemark{a}\\
~~~~$\feh$\dotfill                  &  \hatcurSMEzfeh{} & SPC                 \\
~~~~$\vsini$ (\kms)\dotfill          &  \hatcurSMEvsin{} & SPC                 \\
~~~~$\gamma_{\rm RV}$ (\kms)\dotfill&  \hatcurTRESgamma{} & TRES                  \\
\sidehead{Photometric properties}
~~~~$B$ (mag)\dotfill               &  \hatcurCCtassmB{} & APASS                \\
~~~~$V$ (mag)\dotfill               &  \hatcurCCtassmv{} & APASS               \\
~~~~$I$ (mag)\dotfill               &  \hatcurCCtassmI{} & TASS                \\
~~~~$g$ (mag)\dotfill               &  \hatcurCCtassmg{} & APASS                \\
~~~~$r$ (mag)\dotfill               &  \hatcurCCtassmr{} & APASS                \\
~~~~$i$ (mag)\dotfill               &  \hatcurCCtassmi{} & APASS                \\
~~~~$J$ (mag)\dotfill               &  \hatcurCCtwomassJmag{} & 2MASS           \\
~~~~$H$ (mag)\dotfill               &  \hatcurCCtwomassHmag{} & 2MASS           \\
~~~~$K_s$ (mag)\dotfill             &  \hatcurCCtwomassKmag{} & 2MASS           \\
\sidehead{Derived properties}
~~~~$\mstar$ ($\msun$)\dotfill      &  \hatcurISOmlong{} & Isochrones+\hatcurlumind{}+SPC \tablenotemark{b}\\
~~~~$\rstar$ ($\rsun$)\dotfill      &  \hatcurISOrlong{} & Isochrones+\hatcurlumind{}+SPC         \\
~~~~$\loggstar$ (cgs)\dotfill       &  \hatcurISOlogg{} & Isochrones+\hatcurlumind{}+SPC         \\
~~~~$\lstar$ ($\lsun$)\dotfill      &  \hatcurISOlum{} & Isochrones+\hatcurlumind{}+SPC         \\
~~~~$M_V$ (mag)\dotfill             &  \hatcurISOmv{} & Isochrones+\hatcurlumind{}+SPC         \\
~~~~$M_K$ (mag,\hatcurjhkfilset{})\dotfill&  \hatcurISOMK{} & Isochrones+\hatcurlumind{}+SPC         \\
~~~~Age (Gyr)\dotfill               &  \hatcurISOage{} & Isochrones+\hatcurlumind{}+SPC         \\
~~~~$A_{V}$ (mag) \tablenotemark{c}\dotfill           &  \hatcurXAv{} & Isochrones+\hatcurlumind{}+SPC\\
~~~~Distance (pc)\dotfill           &  \hatcurXdistred{} & Isochrones+\hatcurlumind{}+SPC\\
~~~~$\log{R'_\textrm{HK}}$\dotfill & $-5.0 \pm 0.1$ & Boisse et al 2010\\
\enddata
\tablenotetext{a}{
    SPC = ``Stellar Parameter Classification'' method based on
    cross-correlating high-resolution spectra against synthetic
    templates \citep{buchhave:2012}. These parameters rely primarily
    on SPC, but have a small dependence also on the iterative analysis
    incorporating the isochrone search and global modeling of the
    data, as described in the text.  } 
\tablenotetext{b}{
    Isochrones+\hatcurlumind{}+SPC = Based on the Y$^{2}$ isochrones
    \citep{yi:2001}, the stellar density used as a luminosity indicator, and the 
    SPC results.
} 
\tablenotetext{c}{ Total \band{V} extinction to the star determined
  by comparing the catalog broad-band photometry listed in the table
  to the expected magnitudes from the
  Isochrones+\hatcurlumind{}+SPC model for the star. We use the
  \citet{cardelli:1989} extinction law.  }
\ifthenelse{\boolean{emulateapj}}{
  \end{deluxetable*}
}{
  \end{deluxetable}
}
\ifthenelse{\boolean{emulateapj}}{
  \begin{deluxetable*}{lc}
}{
  \begin{deluxetable}{lc}
}
\tabletypesize{\scriptsize}
\tablecaption{Parameters for the transiting planet \hatcurb{}.\label{tab:planetparam}}
\tablehead{
    \multicolumn{1}{c}{~~~~~~~~Parameter~~~~~~~~} &
    \multicolumn{1}{c}{Value \tablenotemark{a}}                     
}
\startdata
\noalign{\vskip -3pt}
\sidehead{\Lc{} parameters}
~~~$P$ (days)             \dotfill    & $\hatcurLCP{}$              \\
~~~$T_c$ (${\rm BJD}$)    
      \tablenotemark{b}   \dotfill    & $\hatcurLCT{}$              \\
~~~$T_{14}$ (days)
      \tablenotemark{b}   \dotfill    & $\hatcurLCdur{}$            \\
~~~$T_{12} = T_{34}$ (days)
      \tablenotemark{b}   \dotfill    & $\hatcurLCingdur{}$         \\
~~~$\arstar$              \dotfill    & $\hatcurPPar{}$             \\
~~~$\zrstar$ \tablenotemark{c}              \dotfill    & $\hatcurLCzeta{}$\phn       \\
~~~$\rpl/\rstar$          \dotfill    & $\hatcurLCrprstar{}$        \\
~~~$b^2$                  \dotfill    & $\hatcurLCbsq{}$            \\
~~~$b \equiv a \cos i/\rstar$
                          \dotfill    & $\hatcurLCimp{}$           \\
~~~$i$ (deg)              \dotfill    & $\hatcurPPi{}$\phn         \\

\sidehead{Limb-darkening coefficients \tablenotemark{d}}
~~~$c_1,i$ (linear term)  \dotfill    & $\hatcurLBii{}$            \\
~~~$c_2,i$ (quadratic term) \dotfill  & $\hatcurLBiii{}$           \\
~~~$c_1,r$               \dotfill    & $\hatcurLBir{}$             \\
~~~$c_2,r$               \dotfill    & $\hatcurLBiir{}$            \\

\sidehead{RV parameters}
~~~$K$ (\ms)              \dotfill    & $\hatcurRVK{}$\phn\phn      \\
~~~$e$ \tablenotemark{e}  \dotfill    & $\hatcurRVeccentwosiglimemodel{}$ \\
~~~RV jitter NOT~2.6\,m/FIES (\ms) \tablenotemark{f}        \dotfill    & \hatcurRVjitterA{}           \\
~~~RV jitter OHP~1.93\,m/SOPHIE (\ms)        \dotfill    & \hatcurRVjitterB{}           \\

\sidehead{Planetary parameters}
~~~$\mpl$ ($\mjup$)       \dotfill    & $\hatcurPPmlong{}$          \\
~~~$\rpl$ ($\rjup$)       \dotfill    & $\hatcurPPrlong{}$          \\
~~~$C(\mpl,\rpl)$
    \tablenotemark{g}     \dotfill    & $\hatcurPPmrcorr{}$         \\
~~~$\rhopl$ (\gcmc)       \dotfill    & $\hatcurPPrho{}$            \\
~~~$\log g_p$ (cgs)       \dotfill    & $\hatcurPPlogg{}$           \\
~~~$a$ (AU)               \dotfill    & $\hatcurPParel{}$          \\
~~~$T_{\rm eq}$ (K) \tablenotemark{h}        \dotfill   & $\hatcurPPteff{}$           \\
~~~$\Theta$ \tablenotemark{i} \dotfill & $\hatcurPPtheta{}$         \\
~~~$\langle F \rangle$ ($10^{9}$\ergscmsq) \tablenotemark{j}
                          \dotfill    & $\hatcurPPfluxavg{}$       \\ [-1.5ex]
\enddata
\tablenotetext{a}{
    The adopted parameters assume a circular orbit. Based on the
    Bayesian evidence ratio we find that this model is strongly
    preferred over a model in which the eccentricity is allowed to
    vary in the fit. For each parameter we give the median value and
    68.3\% (1$\sigma$) confidence intervals from the posterior
    distribution.
}
\tablenotetext{b}{
    Reported times are in Barycentric Julian Date calculated directly
    from UTC, {\em without} correction for leap seconds.
    \ensuremath{T_c}: Reference epoch of mid transit that
    minimizes the correlation with the orbital period.
    \ensuremath{T_{14}}: total transit duration, time
    between first to last contact;
    \ensuremath{T_{12}=T_{34}}: ingress/egress time, time between first
    and second, or third and fourth contact.
}
\tablenotetext{c}{
    Reciprocal of the half duration of the transit used as a jump
    parameter in our MCMC analysis in place of $\arstar$. It is
    related to $\arstar$ by the expression $\zrstar = \arstar
    (2\pi(1+e\sin \omega))/(P \sqrt{1 - b^{2}}\sqrt{1-e^{2}})$
    \citep{bakos:2010:hat11}.
}
\tablenotetext{d}{
    Values for a quadratic law, adopted from the tabulations by
    \cite{claret:2004} according to the spectroscopic (SPC) parameters
    listed in \reftabl{stellar}.
}
\tablenotetext{e}{
    The 95\% confidence upper-limit on the eccentricity from a model
    in which the eccentricity is allowed to vary in the fit.
}
\tablenotetext{f}{
    Error term, either astrophysical or instrumental in origin, added
    in quadrature to the formal RV errors for the listed
    instrument. This term is varied in the fit assuming a prior inversely 
    proportional to the jitter.
}
\tablenotetext{g}{
    Correlation coefficient between the planetary mass \mpl\ and
    radius \rpl\ determined from the parameter posterior distribution
    via $C(\mpl,\rpl) = <(\mpl - <\mpl>)(\rpl -
    <\rpl>)>/(\sigma_{\mpl}\sigma_{\rpl})>$ where $< \cdot >$ is the
    expectation value operator, and $\sigma_x$ is the standard
    deviation of parameter $x$.
}
\tablenotetext{h}{
    Planet equilibrium temperature averaged over the orbit, calculated
    assuming a Bond albedo of zero, and that flux is reradiated from
    the full planet surface.
}
\tablenotetext{i}{
    The Safronov number is given by $\Theta = \frac{1}{2}(V_{\rm
    esc}/V_{\rm orb})^2 = (a/\rpl)(\mpl / \mstar )$
    \citep[see][]{hansen:2007}.
}
\tablenotetext{j}{
    Incoming flux per unit surface area, averaged over the orbit.
}
\ifthenelse{\boolean{emulateapj}}{
  \end{deluxetable*}
}{
  \end{deluxetable}
}
%


We conducted the analysis inflating the SOPHIE and FIES RV uncertainties 
by adding a "jitter" term in quadrature to the formal uncertainties. This was 
done to accommodate the larger than expected scattering of the RV 
observations around the best-fit model. Independent jitters were used
for each instrument, as it is not clear whether the jitter is instrumental or 
astrophysical in origin. The jitter term was allowed to vary in the fit, yielding a 
$\chi^2$ per degree of freedom of unity for the RVs in the best-fit model. 
The median values for the jitter are \hatcurRVjitterB\,\ms\ for the SOPHIE 
observations and \hatcurRVjitterA\,\ms\ for the FIES observations. This
suggests that either the formal uncertainties of the FIES instrument were
overestimated, or that the jitter from the SOPHIE instrument is not from
the star but from the instrument itself. 

The analysis was done twice: fixing the eccentricity to zero, and allowing
it to vary. Computing the Bayesian evidence for each model, we found 
that the fixed circular model is preferred by a factor of $\sim$500. Therefore
the circular orbit model was adopted. The $95\%$ confidence upper
limit on the eccentricity is $e \hatcurRVeccentwosiglimemodel{}$. 

The best-fit models are presented in Figures \ref{fig:hatnet}, \ref{fig:lc} and 
\ref{fig:rvbis}, and the resulting derived stellar and planetary parameters 
are listed in Tables \ref{tab:stellar} and \ref{tab:planetparam}, respectively. 
We find that the star \hatcur{} has a mass of $\hatcurISOmlong{}$ $\msun$ 
and a radius of $\hatcurISOrlong{}$ $\rsun$, and that its planet \hatcurb{} 
has a period of \hatcurLCP\ days, a mass of $\hatcurPPmlong{}$ 
$\mjup$ and a radius of $\hatcurPPrlong{}$ $\rjup$.


\section{Discussion}
\label{sec:discussion}
We have presented the discovery of a new transiting planet, 
\hatcurb, and provided a precise characterisation of its
properties. \hatcurb\ is a moderately inflated $\sim$0.5 \mjup \ planet, similar in mass, radius and equilibrium 
temperature to HAT-P-1b \citep{bakos:2007:hat1}, WASP-34 \citep{smalley:2011:wasp34}, and
HAT-P-25b \citep{quinn:2012:hat25}. 

With a visual magnitude of V = 13.21, \hatcurb\ is among 
the faintest transiting planet host stars discovered by a wide field ground-based transit  survey 
(today, a total of 11 transiting planet host stars with V $>$ 13 have been discovered by wide-field 
ground based transit surveys, the faintest one is HATS-6 with V = 15.2 \citep{hartman:2014:hats6}). 
Of course, V $>$ 13 is only faint by the standards of surveys like HATNet and WASP; most of the 
hundreds of transiting planets found by space-based surveys such as OGLE, CoRoT 
and Kepler have host stars fainter than \hatcur. It is worth noticing that despite the relative faintness 
of \hatcur, the mass and radius of \hatcurb\ has been measured to better than 10\% precision (relative
to the precision of the stellar parameters) using modest-aperture facilities. This achievement was possible 
because the relatively large size of the planet to its host star provided for a
strong and therefore easy to measure signal. In comparison, only about 140 of all the 1175 known TEP's 
have masses and radii measured to better than 10\% precision. 

\acknowledgements 

\paragraph{Acknowledgements}
HATNet operations have been funded by NASA grants NNG04GN74G and
NNX13AJ15G. Follow-up of HATNet targets has been partially supported
through NSF grant AST-1108686. G.\'A.B., Z.C. and K.P. acknowledge
partial support from NASA grant NNX09AB29G.  K.P. acknowledges support
from NASA grant NNX13AQ62G.  G.T. acknowledges partial support from NASA
grant NNX14AB83G. D.W.L. acknowledges partial support from NASA's Kepler
mission under Cooperative Agreement NNX11AB99A with the Smithsonian
Astrophysical Observatory. D.J. acknowledges the 
Danish National Research Foundation (grant number DNRF97) for partial support.

Data presented in this paper are based on observations obtained at the HAT
station at the Submillimeter Array of SAO, and the HAT station at the
Fred Lawrence Whipple Observatory of SAO. Data are also based on observations 
with the Fred Lawrence Whipple Observatory 1.5m and 1.2m telescopes of SAO. 
This paper presents observations made with the Nordic Optical Telescope, 
operated on the island of La Palma jointly by Denmark, Finland, Iceland,
Norway, and Sweden, in the Spanish Observatorio del Roque de los
Muchachos of the Instituto de Astrof\'isica de Canarias. 
The authors thank all the staff of Haute-Provence Observatory for their
contribution to the success of the ELODIE and SOPHIE projects and their
support at the 1.93-m telescope. The research leading to these results has 
received funding from the European Community's Seventh Framework 
Programme (FP7/2007-2013) under grant agreement number RG226604 
(OPTICON). The authors wish to recognize and acknowledge the very significant cultural role 
and reverence that the summit of Mauna Kea has always had within the
indigenous Hawaiian community. We are most fortunate to have the
opportunity to conduct observations from this mountain.

\bibliographystyle{apj}
\bibliography{htrbib.bib}

\begin{thebibliography}{23}
\expandafter\ifx\csname natexlab\endcsname\relax\def\natexlab#1{#1}\fi

\bibitem[{{Bakos} {et~al.}(2004){Bakos}, {Noyes}, {Kov{\'a}cs}, {Stanek},
  {Sasselov}, \& {Domsa}}]{bakos:2004:hatnet}
{Bakos}, G., {Noyes}, R.~W., {Kov{\'a}cs}, G., {et~al.} 2004, \pasp, 116, 266

\bibitem[{{Bakos} {et~al.}(2007){Bakos}, {Noyes}, {Kov{\'a}cs}, {Latham},
  {Sasselov}, {Torres}, {Fischer}, {Stefanik}, {Sato}, {Johnson}, {P{\'a}l},
  {Marcy}, {Butler}, {Esquerdo}, {Stanek}, {L{\'a}z{\'a}r}, {Papp}, {S{\'a}ri},
  \& {Sip{\H o}cz}}]{bakos:2007:hat1}
{Bakos}, G.~{\'A}., {Noyes}, R.~W., {Kov{\'a}cs}, G., {et~al.} 2007, \apj, 656,
  552

\bibitem[{{Bakos} {et~al.}(2010){Bakos}, {Torres}, {P{\'a}l}, {Hartman},
  {Kov{\'a}cs}, {Noyes}, {Latham}, {Sasselov}, {Sip{\H o}cz}, {Esquerdo},
  {Fischer}, {Johnson}, {Marcy}, {Butler}, {Isaacson}, {Howard}, {Vogt},
  {Kov{\'a}cs}, {Fernandez}, {Mo{\'o}r}, {Stefanik}, {L{\'a}z{\'a}r}, {Papp},
  \& {S{\'a}ri}}]{bakos:2010:hat11}
{Bakos}, G.~{\'A}., {Torres}, G., {P{\'a}l}, A., {et~al.} 2010, \apj, 710, 1724

\bibitem[{{Boisse} {et~al.}(2013){Boisse}, {Hartman}, {Bakos}, {Penev},
  {Csubry}, {B{\'e}ky}, {Latham}, {Bieryla}, {Torres}, {Kov{\'a}cs},
  {Buchhave}, {Hansen}, {Everett}, {Esquerdo}, {Szklen{\'a}r}, {Falco},
  {Shporer}, {Fulton}, {Noyes}, {Stefanik}, {L{\'a}z{\'a}r}, {Papp}, \&
  {S{\'a}ri}}]{boisse:2012:hat4243}
{Boisse}, I., {Hartman}, J.~D., {Bakos}, G.~{\'A}., {et~al.} 2013, \aap, 558,
  A86

\bibitem[{{Bouchy} {et~al.}(2013){Bouchy}, {D{\'{\i}}az}, {H{\'e}brard},
  {Arnold}, {Boisse}, {Delfosse}, {Perruchot}, \& {Santerne}}]{bouchy:2013}
{Bouchy}, F., {D{\'{\i}}az}, R.~F., {H{\'e}brard}, G., {et~al.} 2013, \aap,
  549, A49

\bibitem[{{Buchhave} {et~al.}(2010){Buchhave}, {Bakos}, {Hartman}, {Torres},
  {Kov{\'a}cs}, {Latham}, {Noyes}, {Esquerdo}, {Everett}, {Howard}, {Marcy},
  {Fischer}, {Johnson}, {Andersen}, {F{\H u}r{\'e}sz}, {Perumpilly},
  {Sasselov}, {Stefanik}, {B{\'e}ky}, {L{\'a}z{\'a}r}, {Papp}, \&
  {S{\'a}ri}}]{buchhave:2010:hat16}
{Buchhave}, L.~A., {Bakos}, G.~{\'A}., {Hartman}, J.~D., {et~al.} 2010, \apj,
  720, 1118

\bibitem[{{Buchhave} {et~al.}(2012){Buchhave}, {Latham}, {Johansen},
  {Bizzarro}, {Torres}, {Rowe}, {Batalha}, {Borucki}, {Brugamyer}, {Caldwell},
  {Bryson}, {Ciardi}, {Cochran}, {Endl}, {Esquerdo}, {Ford}, {Geary},
  {Gilliland}, {Hansen}, {Isaacson}, {Laird}, {Lucas}, {Marcy}, {Morse},
  {Robertson}, {Shporer}, {Stefanik}, {Still}, \& {Quinn}}]{buchhave:2012}
{Buchhave}, L.~A., {Latham}, D.~W., {Johansen}, A., {et~al.} 2012, \nat, 486,
  375

\bibitem[{{Cardelli} {et~al.}(1989){Cardelli}, {Clayton}, \&
  {Mathis}}]{cardelli:1989}
{Cardelli}, J.~A., {Clayton}, G.~C., \& {Mathis}, J.~S. 1989, \apj, 345, 245

\bibitem[{{Claret}(2004)}]{claret:2004}
{Claret}, A. 2004, \aap, 428, 1001

\bibitem[{{Djupvik} \& {Andersen}(2010)}]{djupvik:2010}
{Djupvik}, A.~A., \& {Andersen}, J. 2010, in Highlights of Spanish Astrophysics
  V, ed. {J.~M.~Diego, L.~J.~Goicoechea, J.~I.~Gonz{\'a}lez-Serrano, \&
  J.~Gorgas}, 211

\bibitem[{{F\H{u}r\'esz}(2008)}]{furesz:2008}
{F\H{u}r\'esz}, G. 2008, PhD thesis, {Univ. of Szeged, Hungary}

\bibitem[{{Hansen} \& {Barman}(2007)}]{hansen:2007}
{Hansen}, B.~M.~S., \& {Barman}, T. 2007, \apj, 671, 861

\bibitem[{{Hartman} {et~al.}(2012){Hartman}, {Bakos}, {B{\'e}ky}, {Torres},
  {Latham}, {Csubry}, {Penev}, {Shporer}, {Fulton}, {Buchhave}, {Johnson},
  {Howard}, {Marcy}, {Fischer}, {Kov{\'a}cs}, {Noyes}, {Esquerdo}, {Everett},
  {Szklen{\'a}r}, {Quinn}, {Bieryla}, {Knox}, {Hinz}, {Sasselov}, {F{\H
  u}r{\'e}sz}, {Stefanik}, {L{\'a}z{\'a}r}, {Papp}, \&
  {S{\'a}ri}}]{hartman:2012:hat39hat41}
{Hartman}, J.~D., {Bakos}, G.~{\'A}., {B{\'e}ky}, B., {et~al.} 2012, \aj, 144,
  139

\bibitem[{{Hartman} {et~al.}(2014){Hartman}, {Bayliss}, {Brahm}, {Bakos},
  {Mancini}, {Jord{\'a}n}, {Penev}, {Rabus}, {Zhou}, {Butler}, {Espinoza}, {de
  Val-Borro}, {Bhatti}, {Csubry}, {Ciceri}, {Henning}, {Schmidt}, {Arriagada},
  {Shectman}, {Crane}, {Thompson}, {Suc}, {Cs{\'a}k}, {Tan}, {Noyes},
  {L{\'a}z{\'a}r}, {Papp}, \& {S{\'a}ri}}]{hartman:2014:hats6}
{Hartman}, J.~D., {Bayliss}, D., {Brahm}, R., {et~al.} 2014, ArXiv e-prints,
  1408.1758

\bibitem[{{Kov{\'a}cs} {et~al.}(2005){Kov{\'a}cs}, {Bakos}, \&
  {Noyes}}]{kovacs:2005:TFA}
{Kov{\'a}cs}, G., {Bakos}, G., \& {Noyes}, R.~W. 2005, \mnras, 356, 557

\bibitem[{{Kov{\'a}cs} {et~al.}(2002){Kov{\'a}cs}, {Zucker}, \&
  {Mazeh}}]{kovacs:2002:BLS}
{Kov{\'a}cs}, G., {Zucker}, S., \& {Mazeh}, T. 2002, \aap, 391, 369

\bibitem[{{Latham} {et~al.}(2009){Latham}, {Bakos}, {Torres}, {Stefanik},
  {Noyes}, {Kov{\'a}cs}, {P{\'a}l}, {Marcy}, {Fischer}, {Butler}, {Sip{\H
  o}cz}, {Sasselov}, {Esquerdo}, {Vogt}, {Hartman}, {Kov{\'a}cs},
  {L{\'a}z{\'a}r}, {Papp}, \& {S{\'a}ri}}]{latham:2009:hat8}
{Latham}, D.~W., {Bakos}, G.~{\'A}., {Torres}, G., {et~al.} 2009, \apj, 704,
  1107

\bibitem[{{Perruchot} {et~al.}(2011){Perruchot}, {Bouchy}, {Chazelas},
  {D{\'{\i}}az}, {H{\'e}brard}, {Arnaud}, {Arnold}, {Avila}, {Delfosse},
  {Boisse}, {Moreaux}, {Pepe}, {Richaud}, {Santerne}, {Sottile}, \&
  {T{\'e}zier}}]{perruchot:2011}
{Perruchot}, S., {Bouchy}, F., {Chazelas}, B., {et~al.} 2011, in Society of
  Photo-Optical Instrumentation Engineers (SPIE) Conference Series, Vol. 8151,
  Society of Photo-Optical Instrumentation Engineers (SPIE) Conference Series

\bibitem[{{Quinn} {et~al.}(2012){Quinn}, {Bakos}, {Hartman}, {Torres},
  {Kov{\'a}cs}, {Latham}, {Noyes}, {Fischer}, {Johnson}, {Marcy}, {Howard},
  {Szentgyorgyi}, {F{\H u}r{\'e}sz}, {Buchhave}, {B{\'e}ky}, {Sasselov},
  {Stefanik}, {Perumpilly}, {Everett}, {L{\'a}z{\'a}r}, {Papp}, \&
  {S{\'a}ri}}]{quinn:2012:hat25}
{Quinn}, S.~N., {Bakos}, G.~{\'A}., {Hartman}, J., {et~al.} 2012, \apj, 745, 80

\bibitem[{{Smalley} {et~al.}(2011){Smalley}, {Anderson}, {Collier Cameron},
  {Hellier}, {Lendl}, {Maxted}, {Queloz}, {Triaud}, {West}, {Bentley}, {Enoch},
  {Gillon}, {Lister}, {Pepe}, {Pollacco}, {Segransan}, {Smith}, {Southworth},
  {Udry}, {Wheatley}, {Wood}, \& {Bento}}]{smalley:2011:wasp34}
{Smalley}, B., {Anderson}, D.~R., {Collier Cameron}, A., {et~al.} 2011, \aap,
  526, A130

\bibitem[{{Udry} \& {Santos}(2007)}]{udrysantos:2007}
{Udry}, S., \& {Santos}, N.~C. 2007, \araa, 45, 397

\bibitem[{{Wright} {et~al.}(2012){Wright}, {Marcy}, {Howard}, {Johnson},
  {Morton}, \& {Fischer}}]{wright:2012}
{Wright}, J.~T., {Marcy}, G.~W., {Howard}, A.~W., {et~al.} 2012, \apj, 753, 160

\bibitem[{{Yi} {et~al.}(2001){Yi}, {Demarque}, {Kim}, {Lee}, {Ree}, {Lejeune},
  \& {Barnes}}]{yi:2001}
{Yi}, S., {Demarque}, P., {Kim}, Y.-C., {et~al.} 2001, \apjs, 136, 417

\end{thebibliography}

\end{document}